\documentclass[journal]{IEEEtran}

\ifCLASSINFOpdf
\else
   \usepackage[dvips]{graphicx}
\fi
\usepackage{url}

\usepackage{graphicx}
\usepackage{amsmath}
\usepackage{comment}
\usepackage{subcaption}
\usepackage{tikz}
\usetikzlibrary{shapes,arrows}
\usetikzlibrary{external}
\tikzstyle{decision} = [diamond, draw, fill=blue!20, 
    text width=4.5em, text badly centered, node distance=3cm, inner sep=0pt]
\tikzstyle{block} = [rectangle, draw, 
    text width=7em, text centered, minimum height=5em,minimum width=8em,very thick]
\tikzstyle{block2} = [rectangle, draw, fill=red!20, 
    text width=7em, text centered, minimum height=4em,minimum width=8em,very thick]
\tikzstyle{line} = [draw, -latex',very thick]
\tikzstyle{cloud} = [draw, ellipse,fill=red!20, node distance=3cm,
    minimum height=2em]

\begin{document}

\title{Blind Mask to Improve Intelligibility of Non-Stationary Noisy Speech}

\author{F. Farias \IEEEmembership{Student Member, IEEE}, R. Coelho, \IEEEmembership{Senior Member, IEEE}
\thanks{The authors are with the Laboratory of Acoustic Signal Processing, Military Institute of Engineering, Rio de Janeiro, RJ, Brazil. This work was partially supported by the National Council for Scientific and Technological Development (CNPq) under grant 308155/2019-0 and Fundação de Amparo à Pesquisa do Estado do Rio de Janeiro (FAPERJ) under grant 26/203.075/2016. This work is also supported by the Coordenação de Aperfeiçoamento de Pessoal de Nível Superior (CAPES)-Grant Code 001. Email: coelho@ime.eb.br.}}

\markboth{}
{Shell \MakeLowercase{\textit{et al.}}: Bare Demo of IEEEtran.cls for IEEE Journals}
\maketitle

\begin{abstract}
This letter proposes a novel blind acoustic mask (BAM) designed to adaptively detect noise components and preserve target speech segments in time-domain.
A robust standard deviation estimator is applied to the non-stationary noisy speech to identify noise masking elements. 
The main contribution of the proposed solution is the use of this noise statistics to derive an adaptive information to define and select samples with lower noise proportion. Thus, preserving speech intelligibility. 
Additionally, no information of the target speech and noise signals statistics is previously required to this non-ideal mask. The BAM and three competitive methods, Ideal Binary Mask (IBM), Target Binary Mask (TBM), and Non-stationary Noise Estimation for Speech Enhancement (NNESE), are evaluated considering speech signals corrupted by three non-stationary acoustic noises and six values of signal-to-noise ratio (SNR). Results demonstrate that the BAM technique achieves intelligibility gains comparable to ideal masks while maintaining good speech quality.

\end{abstract}


\IEEEpeerreviewmaketitle

\vspace{-1em}
\section{Introduction}

\IEEEPARstart{M}{ost} everyday listening experiences are in the presence of acoustic noise such as car noise, people talking in the background, construction noise, rain and other natural phenomenons. These effects may add unwanted content to a target speech signal while diminish its intelligibility \cite{wang2009speech} and its quality \cite{hu2007comparative}.
Applications such as speaker recognition, speech to text and source localization exhibit lower accuracy when the signal is corrupted by additive noise. 
Thus, the mitigation of this acoustic interference in noisy speech is an important research topic. 
The solutions proposed in the literature are mainly twofold: speech enhancement methods to increase quality, and binary acoustic masks to improve intelligibility.

Speech enhancement schemes mitigate the masking interference to improve the noisy signal quality, usually estimating the noise statistics.
These noise estimation approaches generally consider the frequency or the time domain. 
Methods in the frequency domain, such as Spectral Subtraction (SS) \cite{boll1979suppression} and the Optimally Modified Log-Spectral Amplitude (OMLSA) \cite{cohen2002optimal} use some transform to represent signals in the frequency domain and then estimate the noise components. 
Methods in the time domain usually estimate noise statistics with statistical estimators as the present in Non-stationary Noise Estimation for Speech Enhancement (NNESE) \cite{tavares2016speech} or time-frequency decomposition, e.g. in the EMD-Based filtering with Hurst exponent (EMDH) \cite{zao2014speech,coelho2015empirical}.
Although speech enhancement algorithms successfully improve speech quality, they are not designed to achieve intelligibility gain. 
This is particularly challenging in non-stationary environments.
In some cases, the suppression of noisy components causes a distortion that hinders intelligibility \cite{loizou2011reasons}.

Acoustic masks \cite{li2009optimality,kim2010improving,zao2014time} are developed to reduce the perceptual effects of noisy speech to increase intelligibility of a variety of applications.
These methods are devised to emulate the capacity of the human auditory system to segregate a specific sound even in the presence of many others, also known as the \textit{cocktail party} effect \cite{wang2005ideal}. 
Acoustic binary masks can be classified as ideal or blind. 
Ideal masks are constructed using information of the clean speech, as well as the noisy speech.
The Ideal Binary Mask (IBM) \cite{brown1994computational} is built comparing the energy of the signal and of the noise in each Time-Frequency (T-F) region. 
The Target Binary Mask (TBM) \cite{anzalone2006determination} builds its mask comparing the energy of the clean signal with the Speech Shaped Noise (SSN).
Blind masks are made based on an estimation of clean speech characteristics from the noisy signal \cite{yilmaz2004blind,hazrati2013blind}. 
However, the accuracy of this estimation usually depends on extensive training using neural networks and large databases.
Binary masked speech may present high objective quality scores,
but the abrupt difference between the retained and discarded regions of a binary masked signal often cause musical noise, which can lead to quality loss\cite{li2008factors}.

This letter proposes a blind acoustic mask in the time domain to improve speech intelligibility.
The main idea of this strategy is to estimate noise components and the proportion of the speech signal in each short-time frame.
This information is used to delimit a set of samples proportional to the presence of the target signal in each frame, so if the frame is mostly comprised by the target signal, these samples take most of the frame. While the frame is processed to mitigate the effects of noise, the samples in the delimited set are preserved, thus maintaining speech intelligibility.
Additionally, as a blind mask it avoids the usage of previous information from the clean speech and noise.

Several experiments are conducted to evaluate the proposed mask in terms of speech intelligibility and quality. 
The noisy scenario is composed by three background acoustic noises with six different SNR values. 
Three objective measures are adopted for intelligibility evaluation. While Short Time Objective Intelligibility (STOI) \cite{taal2011algorithm} is the state-of-the-art in intelligibility prediction, the Approximated Short-Time Speech Intelligibility Index (ASII) \cite{taal2013optimal} and Extended Speech Intelligibility Index (ESII) \cite{rhebergen2005speech} are designed to deal with non-stationary distortions.
Objective quality evaluation is also performed using Perceptual Evaluation of Speech Quality (PESQ) \cite{recommendation2001perceptual} measure.
The Index of Non-Stationarity (INS) \cite{borgnat2010testing} is also selected to analyse the effect of the proposed mask and the competing methods.

\section{BAM: Blind Acoustic Mask}

\begin{figure}[t]
    \centering
    \includegraphics[width=0.7\columnwidth]{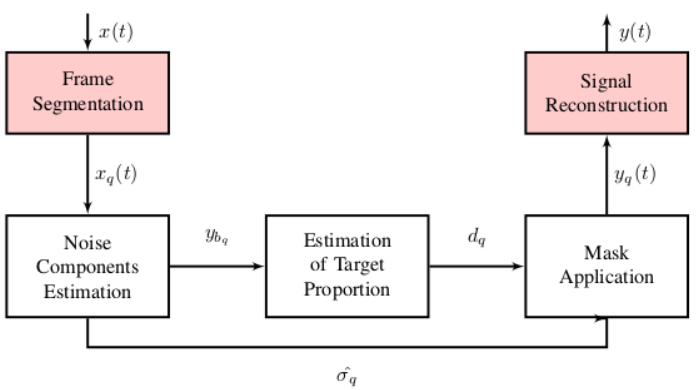}
    \caption{Schematic of the proposed Blind Acoustic Mask.}
    \label{fig:schema}
    \vspace{-1.5em}
\end{figure}

The schematic of the proposed blind acoustic mask is depicted in Figure \ref{fig:schema}. It consists of three main steps: 
First, the signal is separated into non-overlapping short-time frames. In each frame, the noise standard deviation is detected using a robust estimator. 
In the second step, this information is used to derive the parameter $d_q$ that refers to the proportion of speech that is present in the $q$-th frame.
Then the adaptive mask is applied in each frame, according to the parameter $d_q$, and the processed frames are concatenated to reconstruct the processed signal. 
This mask employs the noise statistics estimation used in \cite{tavares2016speech}.
The aim is to improve intelligibility of speech corrupted by additive noise, while maintaining the quality gain obtained using a speech enhancement technique.

\vspace{-1em}
\subsection{Step 1: Noise Components Estimation}

This step begins with the segmentation of the signal in non-overlapping short-time frames. In each frame, the $d$-Dimensional Trimmed Estimator (DATE) \cite{pastor2012robust} is used to detect the noise standard deviation. This estimator was first defined to work on signals mixed with Gaussian noise over the entire signal. 
However, as shown in \cite{tavares2016speech} it also works with different noise distributions.

First, the samples $x_q(t)$ from the $q$-th frame are organized from lower to higher amplitude $Y_1 \leq Y_2 \leq ... \leq Y_T$.
Then a value $t_{min}$ is computed, such as the sample amplitudes lower than $Y_{t_{min}}$ are considered as only noise.
The value of $b_q$ is the lowest value of $t$ that is higher than $t_{min}$ and obeys the relation $ ||Y_{t-1}|| \leq \frac{c \sum_{i=1}^T ||Y_i||}{t} \leq ||Y_{t+1}||$, where $c$ is an adjustment factor that depends on the detection threshold.
The noise standard deviation for that frame is then estimated as

\vspace{-1em}
\begin{equation}
	\hat{\sigma_q} = \frac{c \sum_{i=1}^{b_q} ||Y_i||}{b_q}.
	\label{eq:nnese}
\end{equation}

In this step it is also defined the value $y_{b_q}$, amplitude from the vector $Y_i$ associated to the value $b_q$. Any sample lower than  $y_{b_q}$ is considered as noise.

\vspace{-1em}
\subsection{Step 2: Estimation of Target Proportion $d_q$}

The removal of samples with amplitude values below $y_{b_q}$ may yield an improvement in quality.
However, some of those frames are mainly composed by the target signal. The modification of samples from those frames usually compromises intelligibility.  
Thus, a parameter $d_q$ is defined to identify frames where the target signal is prevalent:

\begin{equation}
    d_q = \frac{|\sigma_{q_{ny}} - \hat{\sigma_q}|}{|\sigma_{q_{ny}} + \hat{\sigma_q}|}
\end{equation}

\noindent where $\hat{\sigma_q}$ is the estimated noise standard deviation of the frame $q$ and $\sigma_{q_{ny}}$ is the noisy signal standard deviation.

\begin{figure}[t]
    \centering
    \begin{subfigure}{.9\linewidth}
    \vspace{-1em}
    \centering
    \includegraphics[width=\linewidth]{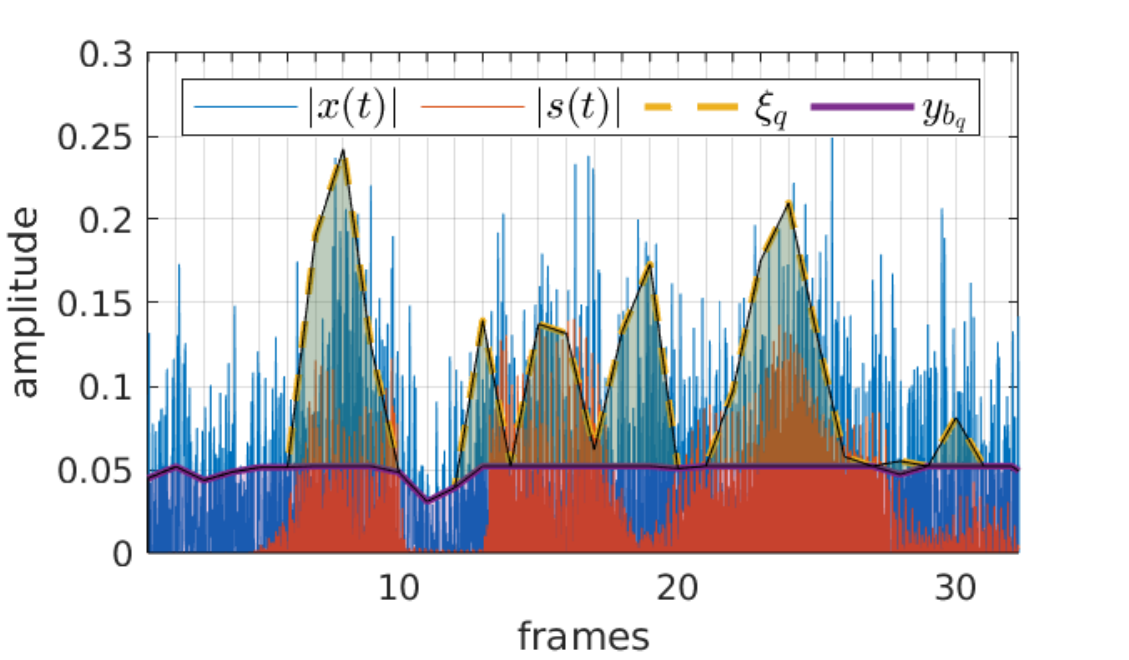}
    \end{subfigure}
    \caption{Clean speech signal $|s(t)|$, corresponding speech signal corrupted with Babble noise at SNR = -6 dB $|x(t)|$, lower threshold $y_{bq}$ and upper threshold $\xi_q$.}
    
    \label{fig:mask_nne}
    \vspace{-1.5em}
\end{figure}

\vspace{-1em}
\subsection{Step 3: Mask Application}

The proposed mask defines which samples are left unaltered in each frame.
These samples correspond to a region where $d_q$ is greater than the lower amplitude $y_{b_q}$.
The lower bound of this region is defined by $y_{b_q}$ and the upper bound is defined through an adaptive threshold $\xi_q$.

\vspace{-1em}
\begin{equation}
    \xi_q = max(y_{b_q},d_q).
\end{equation}

Each sample of the $q$-th frame of the processed signal is then given by

\vspace{-1em}
\begin{equation}
   y_q(t)=\begin{cases}
    x_q(t),& \text{if } y_{b_q} < x_q(t) < \xi_q ;\\
    x_q(t)-\alpha \cdot \hat{\sigma_q},& \text{if } x_q(t) \geq \xi_q;\\
    \beta \cdot x_q(t),              & \text{otherwise}.
\end{cases}
\label{eq:processings}
\end{equation}

\noindent where $\alpha$ is the over-subtraction factor for the speech signal reconstruction and $\beta$ is the flooring factor for negative amplitude values. The frames are then concatenated to form the processed signal $y(t)$.

\begin{figure*}[t]
    \centering
    \begin{subfigure}{.2\textwidth}
        \centering
        \includegraphics[width=\linewidth]{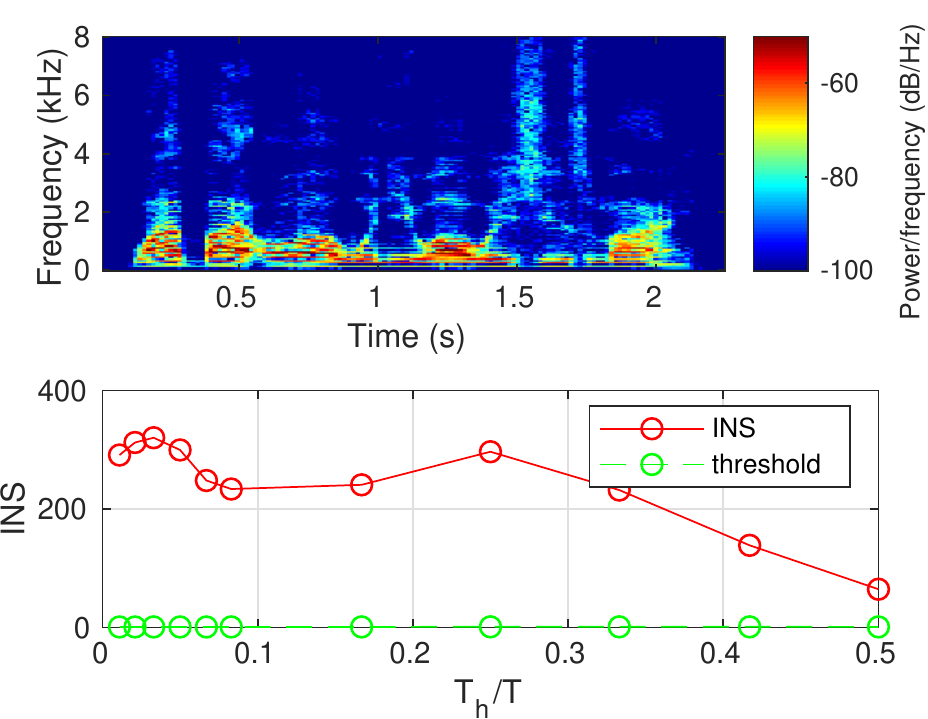}
        \caption{Clean}
    \end{subfigure}%
    \begin{subfigure}{.2\textwidth}
        \centering
        \includegraphics[width=\linewidth]{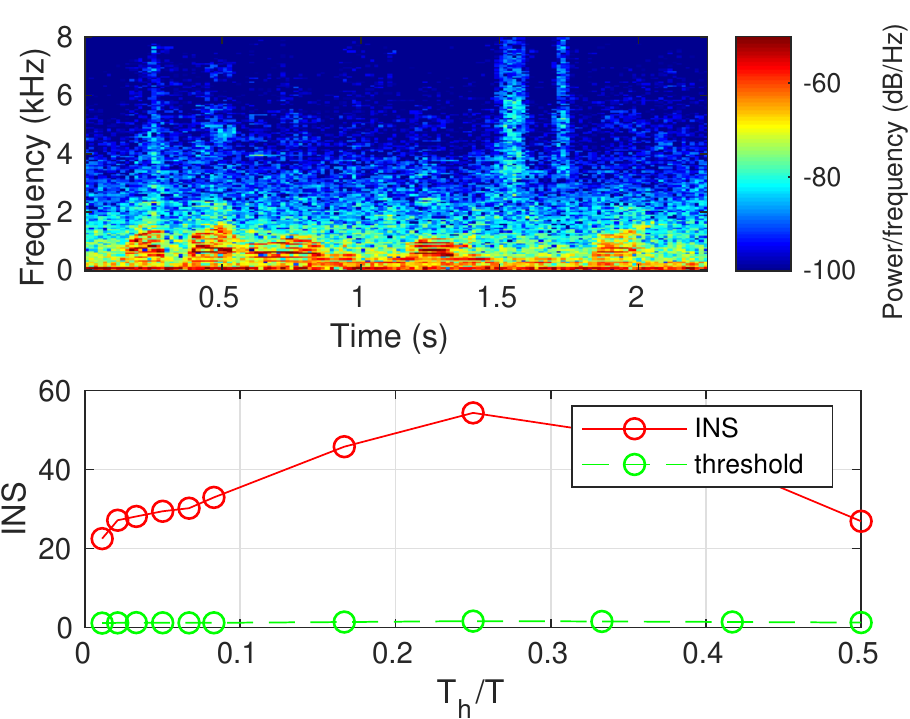}
        \caption{Noisy}
    \end{subfigure}%
    \begin{subfigure}{.2\textwidth}
        \centering
        \includegraphics[width=\linewidth]{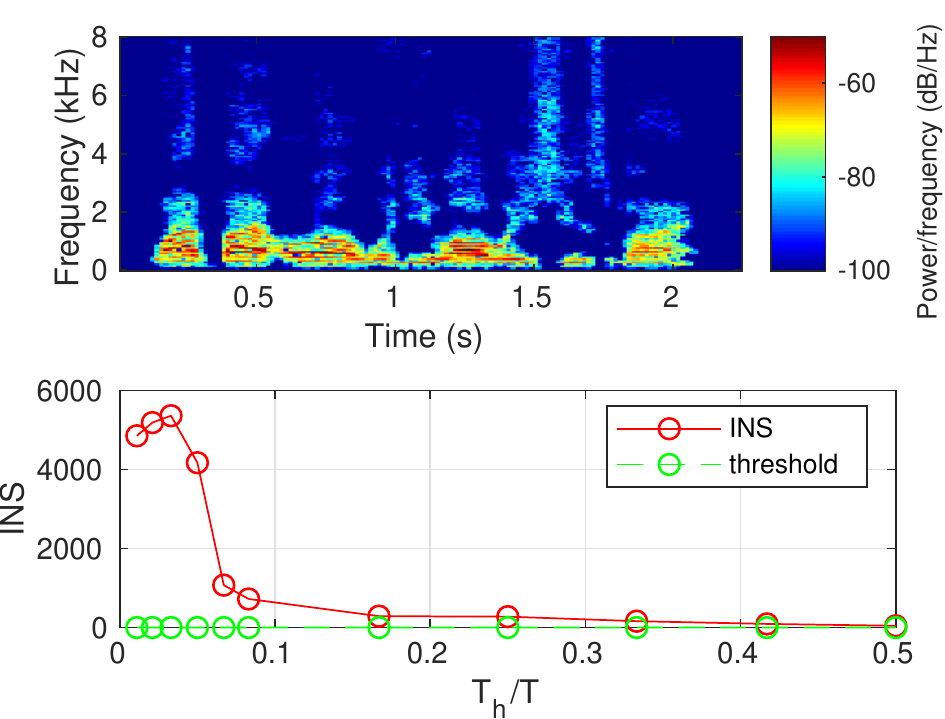}
        \caption{IBM}
    \end{subfigure}%
    \begin{subfigure}{.2\textwidth}
        \centering
        \includegraphics[width=\linewidth]{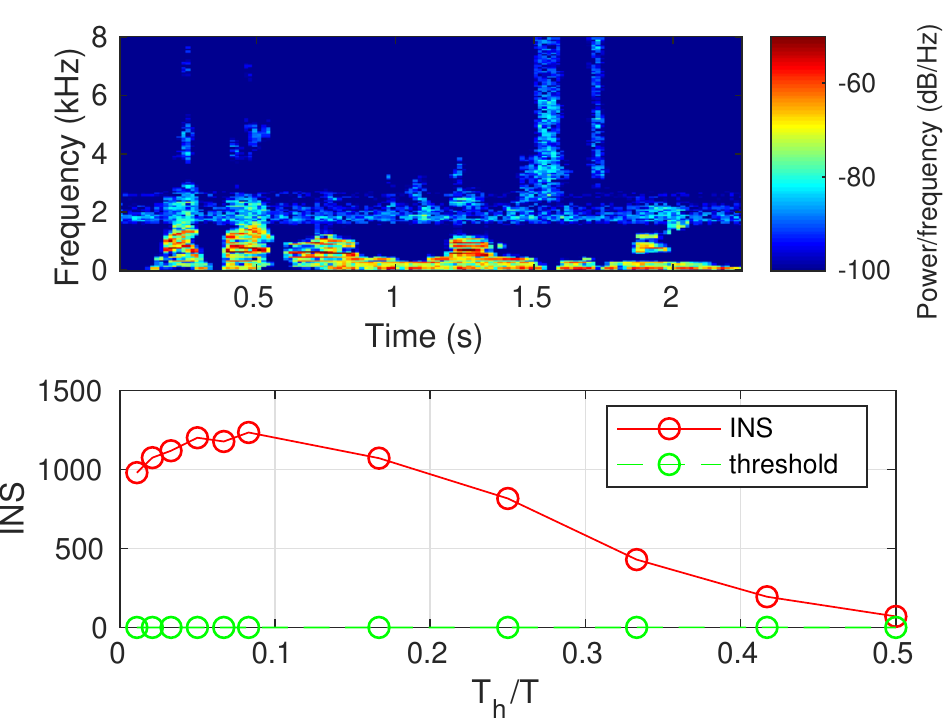}
        \caption{TBM}
    \end{subfigure}%
    \begin{subfigure}{.2\textwidth}
        \centering
        \includegraphics[width=\linewidth]{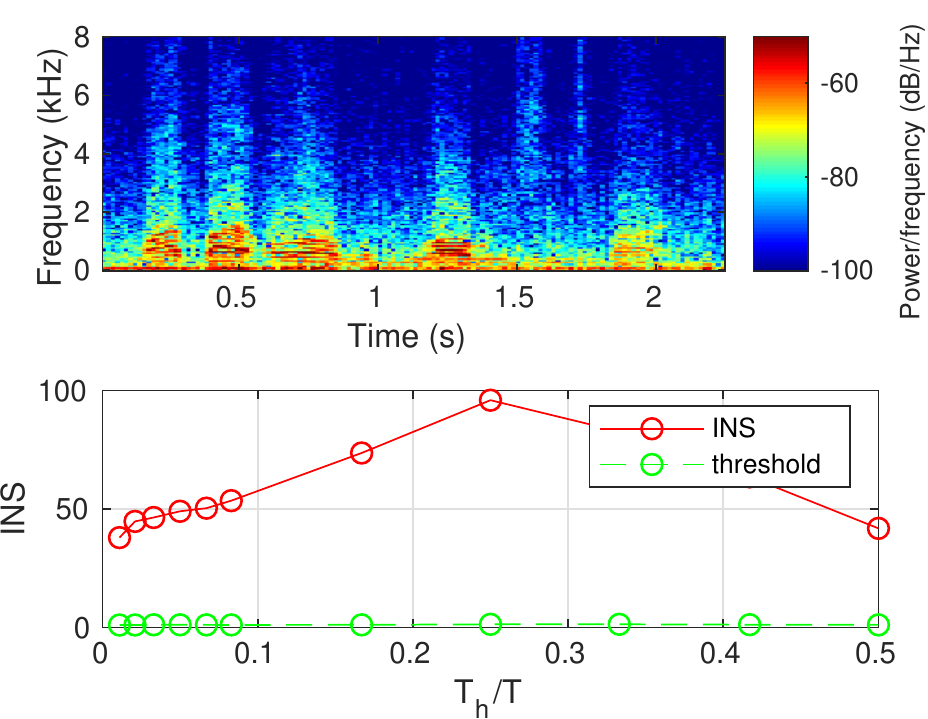}
        \caption{BAM}
    \end{subfigure}%

    \caption{Spectrograms and INS of (a) clean speech, (b) unprocessed speech corrupted with Factory noise at SNR = 3 dB, noisy speech processed with (c) IBM, (d) TBM and (e) the proposed BAM.}
    \vspace{-1.5em}
    \label{fig:specs}
\end{figure*}

The region in each frame is illustrated in Figure \ref{fig:mask_nne}, 
which depicts the clean signal $s(t)$, the noisy signal $x(t)$, the lower bound $y_{b_q}$ and upper threshold $\xi_q$ of the masked region for each frame $q$.
Note that frames 23-27 are mainly composed by speech. Thus, the mask preserves most of the signal in these frames.

\section{Experiments and Discussion}

\begin{table}[t]
\centering

\caption{Intelligibility measures for UNP speech signals.}
\resizebox{.85\columnwidth}{!}{
\begin{tabular}{c|cccccc}
\hline
\textbf{}          & \multicolumn{6}{c}{STOI}                                                  \\
\hline
SNR (dB)        & \textbf{-6} & \textbf{-5} & \textbf{-3} & \textbf{0} & \textbf{3} & \textbf{5}     \\
\hline
\textbf{Babble}     & 0.355 & 0.356 & 0.388 & 0.447 & 0.499 & 0.529 \\ 
\textbf{Cafeteria}  & 0.357 & 0.381 & 0.419 & 0.458 & 0.503 & 0.540 \\ 
\textbf{Factory}    & 0.436 & 0.476 & 0.506 & 0.557 & 0.601 & 0.654 \\ 
\hline

\hline
\textbf{}          & \multicolumn{6}{c}{ASII$_{ST}$}                                                  \\
\hline
SNR (dB)        & \textbf{-6} & \textbf{-5} & \textbf{-3} & \textbf{0} & \textbf{3} & \textbf{5}     \\
\hline
\textbf{Babble}  & 0.348 & 0.365 & 0.391 & 0.433 & 0.497 & 0.519 \\ 
\textbf{Cafeteria} & 0.364& 0.377& 0.410& 0.446& 0.504& 0.523 \\ 
\textbf{Factory} & 0.398& 0.415& 0.446& 0.501& 0.555& 0.586 \\ 
\hline

\hline
\textbf{}          & \multicolumn{6}{c}{ESII}                                                  \\
\hline
SNR (dB)        & \textbf{-6} & \textbf{-5} & \textbf{-3} & \textbf{0} & \textbf{3} & \textbf{5}     \\
\hline

\textbf{Babble}    & 0.306& 0.329& 0.361& 0.416& 0.497& 0.525 \\ 
\textbf{Cafeteria}  & 0.327& 0.344& 0.386& 0.432& 0.505& 0.528 \\ 
\textbf{Factory} & 0.371& 0.394& 0.433& 0.503& 0.570& 0.608 \\ 
\hline
\end{tabular}
}

\label{tab:unp}
\end{table}

The proposed technique is evaluated in terms of intelligibility and quality considering several noisy conditions. It is compared to baseline speech enhancement NNESE \cite{tavares2016speech}, and acoustic masks TBM \cite{anzalone2006determination} and IBM \cite{wang2005ideal}. 
The speech enhancement is considered the baseline for quality gain, as the IBM for intelligibility improvement. 
For the objective evaluation, a subset of 20 utterances from the TIMIT database \cite{garofolo1993darpa} is randomly selected to compose each scenario, leading to 120 tests per method. 
Each segment is sampled at 16 kHz and has average time duration of 3 seconds. 
The acoustic noises Babble and Factory were selected from the RSG-10 \cite{steeneken1988description} database while Cafeteria noise was selected from DEMAND \cite{thiemann2013demand} database.

Speech signals are corrupted considering six SNRs: -6 dB, -5 dB, -3 dB, 0 dB, 3 dB and 5 dB. 
NNESE and the proposed mask are applied on a 32 ms frame-by-frame basis.
Parameters $\alpha$ and $\beta$ are set to $\alpha= 0.35$ and $\beta = 0.65$.
IBM and TBM separate signals in 20 ms Time-Frequency regions, with 10 ms overlapping. 
Frequency separation is performed through a 64-channel gammatone filterbank with center frequencies ranging from 50 to 8000 Hz according to the Equivalent Rectangular Bandwidth\cite{patterson1987efficient}. 
The IBM Relative Criterion is set to -5 dB, as recommended in \cite{kjems2009role}. The TBM is set to detect $99\%$ of the speech energy in each frame.

The intelligibility scores of the unprocessed signals are presented in Table \ref{tab:unp}. The SNR values were chosen such as the STOI score of the UNP signals vary between 0.45 and 0.75, which are considered the threshold of poor and good intelligibility, respectively \cite{sauert2006near}.

Table \ref{tab:proctime} indicates the computational complexity, here represented by the processing time required for each method evaluated for 512 samples per frame. 
These values are normalized by the execution time of the proposed BAM. 
Note that the proposed mask is almost 10 times faster than the binary masks.

\begin{table}[t]
    \centering
    \caption{Normalized Mean Processing Time.}
    \begin{tabular}{c|c|c|c}
        \hline
        NNESE & IBM & TBM & BAM\\
         \hline
        0.5 & 9.4 & 9.2 & 1.0 \\
         \hline
    \end{tabular}
    
    \vspace{-1.5em}
    \label{tab:proctime}
\end{table}

\vspace{-1em}
\subsection{Index of Non-Stationarity}

The Index of Non-Stationarity (INS) \cite{borgnat2010testing} is here adopted to objectively evaluate the non-stationarity of noisy speech signals.
This measure is obtained comparing the target signal with a set of stationary references called \textit{surrogates} 
at different time scales $T_h/T$, where $T_h$ is a short-time analysis window and $T$ is the total duration of the signal. 
The surrogates are obtained from the signal, maintaining the magnitude of the data spectrum and replacing the phase by a random sequence uniformly distributed. 
A threshold $\gamma $ is defined for each window length $T_h$, considering a confidence degree of 95\%.

\vspace{-1em}
\begin{equation*}
    INS \begin{cases} 
            \leq \gamma, \text{signal is stationary};\\
            > \gamma, \text{signal is non-stationary}.
        \end{cases}
\end{equation*}

Figure \ref{fig:specs} depicts spectrograms and INS values for a speech signal, the corresponding signal with Factory noise at 3 dB, the noisy signal processed by the IBM, TBM, and the proposed BAM.
Note that the noise modifies the temporal and spectral characteristics of the clean signal.
The non-stationary behavior of the speech is also softened by noise. The maximum INS ($INS_{max}$) changes from 320 in speech signal to 55 in noisy speech. 
The proposed BAM restores some of the spectral and temporal characteristics of the signal. This can be seen near 0.5 s, where the separation between formants is lost by the noise and recovered processing with BAM. Additionally, the proposed mask restores some of the non-stationary behavior of the signal, the $INS_{max}$ value is increased to 95.
Although the TBM and IBM repair some of the spectral characteristics of the speech signal, they significantly increase the non-stationarity degree. The $INS_{max}$ reaches values above 5000 using the IBM and above 1200 using the TBM.
The $INS_{max}$ values achieved for IBM and TBM are explained by its binary effect particularly for zeroing masking condition.
Table \ref{tab:ins_noises} presents the maximum INS for speech with three studied additive noises. The effect of the proposed BAM depicted in Figure \ref{fig:specs} is also present in the Cafeteria noise, while in Babble noise there is a slight decrease in INS.

\begin{table}[t]
    \centering
    \caption{$INS_{max}$ for UNP and masked signal.}
    
    \begin{tabular}{c|c|c|c|c}
        \hline
        noise & UNP & IBM & TBM & BAM\\
         \hline
        Babble & 80.2 & 6200.4 & 1280.6 & 64.2 \\
        Cafeteria & 35.3 & 5138.3 & 1290.7 & 56.0 \\
        Factory & 54.4 & 5368.8 & 1235.0 & 96.0 \\
         \hline
    \end{tabular}
    
    \vspace{-1.5em}
    \label{tab:ins_noises}
\end{table}

\begin{figure}[t]
    \centering
    \resizebox{.5\textwidth}{!}{
    \includegraphics[width=\linewidth]{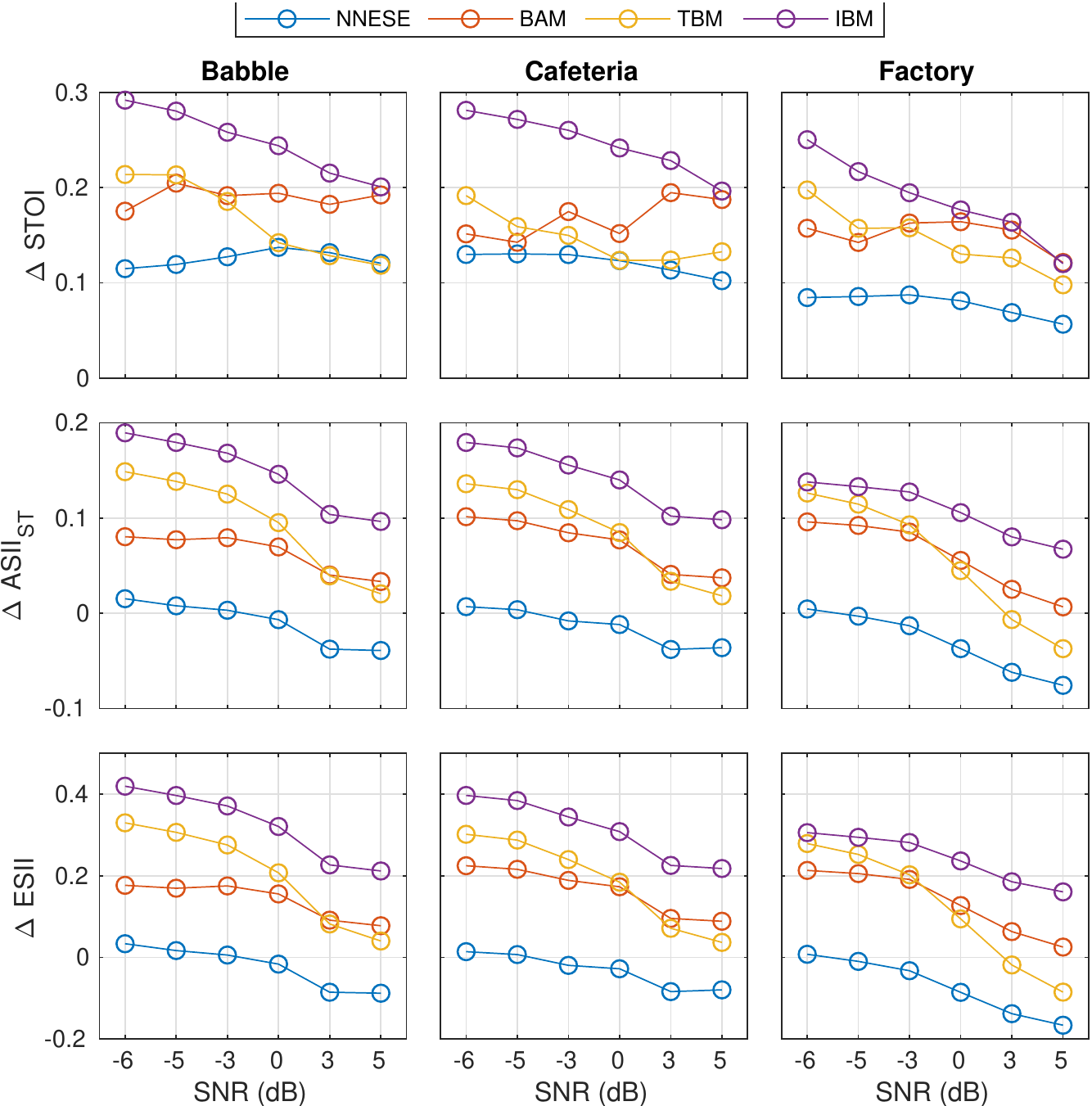}
    }
        \caption{Average improvement in intelligibility for noisy speech with Babble, Cafeteria and Factory.}
        
        \vspace{-1.5em}
        \label{fig:improv_obj}
        
\end{figure}

\subsection{Objective Intelligibility Evaluation}

Three objective measures are adopted to evaluate intelligibility gain of the proposed mask and the competitive methods.
STOI \cite{taal2011algorithm} is a measure developed to predict the intelligibility of noisy speech processed by T-F weighting masks, such as speech enhancement methods.
The metric is the correlation coefficient between the spectral envelopes of clean and enhanced signal.
ASII$_{ST}$ \cite{taal2013optimal} and ESII \cite{rhebergen2005speech} are based on the classic Speech Intelligibility Index (SII) \cite{american1997american}, designed to deal with the non-stationarity of speech and its distortions. 
The ESII score is the average of the SII in each short-time frame of the signal. ASII$_{ST}$ is an adaptation of the ESII, modelling intelligibility after a function of the SII. Both measures are based on the weighted SNR of the signal.
All measures vary between 0 and 1, in which 1 represents a fully intelligible sentence.
The STOI objective measure is normalized by the intelligibility achieved for the target signal corrupted by SSN noise at 10 dB, considered here as a good intelligibility reference.

The average improvement in intelligibility is presented in Figure \ref{fig:improv_obj}. 
Considering $\Delta$STOI, in the top row figures, 
as the baseline for speech intelligibility, IBM presents the highest average intelligibility improvement, 0.25 for Babble noise, followed by 0.19 for the proposed BAM and 0.17 for the proposed TBM. 
While in Cafeteria noise the improvement is similar to the obtained in Babble, in Factory noise the improvement by all methods is lower (0.19 by IBM, 0.15 by BAM and 0.14 by TBM). 
Though it can be noticed that speech with Factory noise achieves the highest UNP intelligibility.

The improvement obtained by ASII$_{ST}$ measure ($\Delta$ASII$_{ST}$) is shown in the middle row. 
It can be seen that the proposed mask improves intelligibility in almost every condition, but this effect is accentuated in lower SNRs. The average ASII$_{ST}$ improvement of the mask considering SNR = -3 dB is 0.08, while the improvement with SNR = 3 dB is 0.02.
It is also noted that the proposed mask achieves superior gain when compared to TBM for SNR values above 0 dB.

The ESII improvement ($\Delta$ESII) is depicted in the bottom row figures. Similarly to $\Delta$ASII$_{ST}$ results, the proposed BAM shows interesting improvement for SNR values lower than 3 dB. 
Maximum $\Delta$ESII is 0.08 for the Babble noise at SNR = -6 dB, 0.10 for Cafeteria noise at SNR = -6 dB and 0.09 in Factory noise at the same SNR. 
The average improvement is comparable to the achieved by the TBM.

\begin{figure}[t]
    \centering
    \resizebox{.5\textwidth}{!}{
    \includegraphics[width=1\linewidth]{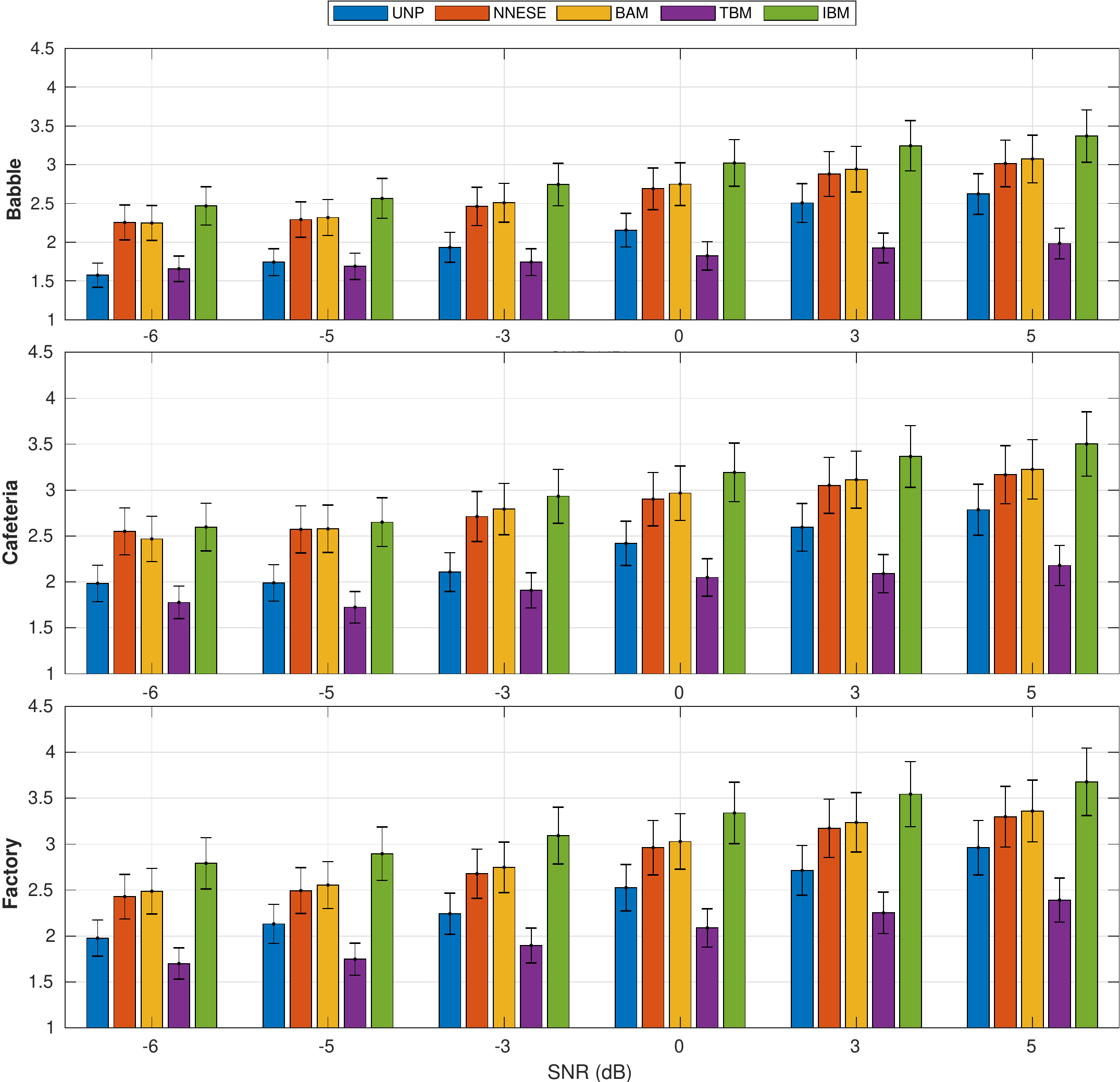}
   }
    \caption{Average PESQ score for noisy speech with Babble, Cafeteria and Factory.}
    \vspace{-1.5em}
    \label{fig:pesq}
\end{figure}

\vspace{-1em}
\subsection{Objective Quality Evaluation}

Speech quality is evaluated using the objective PESQ measure \cite{recommendation2001perceptual}, developed to assess quality of narrow-banded speech and handset telephony. The PESQ score varies from -0.5 to 4.5, in which 4.5 is the maximum quality. 
A signal is considered of fair quality when its PESQ score is above 2, given this objective metric aims to predict Mean Opinion Scores (MOS) \cite{rothauser1969ieee}.

PESQ scores are presented in Figure \ref{fig:pesq}. 
The best quality improvement is obtained using the IBM in all SNR conditions. The BAM presents the highest improvement among the remaining techniques. In the Babble noise, the proposed mask achieves an average PESQ score of 2.64, against 2.60 of the NNESE and 1.80 of the TBM. In the Cafeteria noise, the proposed mask achieves an average PESQ score of 3.10, against 3.04 of the NNESE and 2.11 of the TBM. In the Factory noise, score achieved by the mask is of 2.90, against 2.84 of the NNESE and 2.01 of the TBM.

\section{Conclusion}

This letter introduced a time domain blind acoustic mask to improve intelligibility of speech signals corrupted by non-stationary noises with different INS and SNR values. The proposed mask is evaluated considering the STOI, ASII$_{ST}$, ESII, and PESQ objective measures. Results demonstrate that the proposed BAM outperforms baseline masks with an overall gain of $21.7\%$ in terms of speech intelligibility while keeping good quality gain when compared to the speech enhancement competitive approach.


\bibliographystyle{IEEEtran}
\bibliography{refs}

\end{document}